\documentclass[twocolumn]{article}
\pdfoutput=1

\usepackage{upgreek}
\usepackage{authblk}
\usepackage{geometry}
\usepackage{cite}
\usepackage{hyperref}
\usepackage{fancyhdr}
\usepackage[toc,page]{appendix}
\usepackage{pdfpages}
\usepackage{todonotes}
\usepackage{subfigure}
\usepackage{bigints}
\usepackage{siunitx}
\usepackage{graphicx}
\usepackage{amssymb}
\usepackage{amsmath}
\usepackage[infoshow,debugshow]{tabularx}
\usepackage{relsize}
\usepackage{color}
\usepackage{cleveref}
\usepackage{indentfirst}
\usepackage{setspace}
\usepackage[english]{babel}
\usepackage{mathtools}
\usepackage{nicefrac}

\newcommand{\three}{\SI{300}{\radian \, \metre^{-1}\second^{-1}}}
\DeclareMathOperator{\T}{T}
\DeclareMathOperator{\tr}{tr}
\DeclareMathOperator{\de}{d}

\begin{document}

\title{\textbf{Poynting effect of brain matter in torsion}}

\author[1]{Valentina Balbi}
\author[2]{Antonia Trotta}
\author[1,2]{Michel Destrade}
\author[2]{Aisling N\'i Annaidh}
\affil[1]{\small{School of Mathematics, Statistics and Applied Mathematics, NUI Galway,  University Road, Galway, Ireland}}
\affil[2]{\small{School of Mechanical \& Materials Engineering, University College Dublin, Belfield, Dublin 4}}

\date{}


\twocolumn[
  \begin{@twocolumnfalse}
    \maketitle
   \begin{abstract}

We investigate experimentally and model theoretically the mechanical behaviour of brain matter in torsion. 
Using a strain-controlled rheometer we perform torsion tests on fresh porcine brain samples. 
We quantify the torque and the normal force required to twist a cylindrical sample at constant twist rate. 
Data fitting gives a mean value for the shear modulus $\mu=900\pm 312\, \si{\pascal}$ and for the second Mooney-Rivlin parameter $c_2=297\pm 189\,\si{\pascal}$, indicative of extreme softness. 
Our results show that brain always displays a positive Poynting effect; in other words, it expands in the direction perpendicular to the plane of twisting. 
We validate the experiments with Finite Element simulations and show that when a human head experiences a twisting motion in the horizontal plane, the brain can experience large forces in the axial direction.

\end{abstract}
\vspace{3em}

  \end{@twocolumnfalse}
]


\section{Introduction}


The brain is an extremely soft and fragile tissue, making it hard to test experimentally using the standard protocols in place for soft matter such as elastomers or rubbers\cite{BS12}.
The inflation test is not appropriate for obvious reasons.
The uni-axial tensile test requires a dogbone geometry and clamping, which cannot be realised in practice for brain tissue; instead, the end faces of a cylinder have to be glued to the tension plates, which leads rapidly to an inhomogeneous deformation\cite{miller2002mechanical, rashid2012inhomogeneous}.
The compression test can achieve homogeneous deformation with lubrication of the plates, but only up to about 10\% strain, after which it starts to bulge out\cite{rashid2012mechanicalb}.
By contrast, the simple shear test works well\cite{donnelly1997shear,destrade2015extreme} to least 45$^\circ$ tilting angle leading to a maximal stretch of more than 60\%.  
Similarly, as we show in this paper, the torsion test can be implemented readily for brain matter.

\begin{figure}[b!]
\centering
\includegraphics[width=0.46\textwidth]{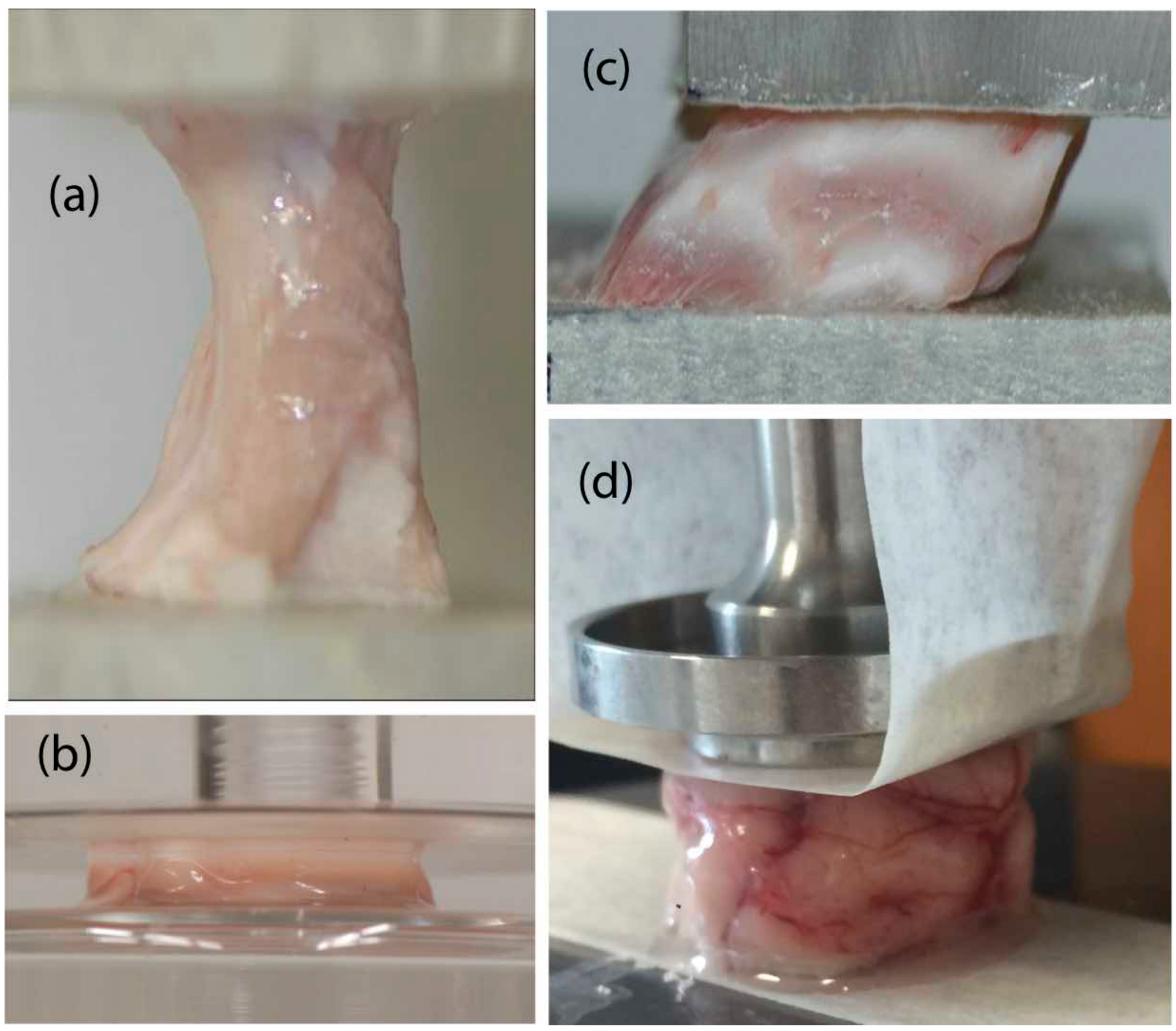}
\caption{Standard testing protocols for soft solids applied to brain matter (porcine): (a) tensile test with glued ends and (b) compression test with lubricated faces: notice the inhomogeneity of the resulting deformations; (c) simple shear and (d) torsion: here the samples behave as required.}
\label{fig:protocols}
\end{figure}

Simple shear and torsion tests are particularly useful to study the Poynting effect, a typical nonlinear phenomenon displayed by soft solids. 
When sheared or twisted those materials tend to elongate (positive Poynting effect\cite{rivlin1949large}) or contract (negative Poynting effect\cite{janmey2007negative}) in the direction perpendicular to the shearing or twisting plane. 
This phenomenon has been observed for brain matter in simple shear tests\cite{destrade2015extreme}. 
However, a practical limitation of simple shear tests is that to date there are no shearing devices able to measure and quantify the normal force, which limits the determination of material parameters.\\
\indent An alternative test is torsion.
It can be performed by glueing a cylindrical sample between two parallel plates and then applying a twist to the sample by rotating one plate with respect to the other. 
In the past, torsion tests on brain matter have been performed using a \textit{rheometer}, at constant strain rates (from $0.05$ to $1\si{\second^{-1}}$)\cite{bilston2001large} to measure the elastic properties of the tissue and dynamically over a range of frequencies ($20$-$200$ \si{\hertz})\cite{arbogast1998material} to investigate its viscoelastic behaviour. 
A comprehensive summary of the results of mechanical tests on brain tissue can be found in two recent reviews\cite{chatelin2010fifty,goriely2015mechanics}. 
However, in those studies only the torques were recorded. 
Moreover, torsion was modelled as simple shear, an equivalence which as we show in Section 5, is only valid locally. 
To the authors' knowledge, the role played by the normal forces arising during torsion has not been investigated yet. \\
\indent In this work, we perform torsion tests on cylindrical porcine brain samples and measure the torque and the axial force required to twist the samples at a constant twist rate of $\SI{300}{\radian\, \meter^{-1}\second^{-1}}$.  
In Section \ref{matmet} we describe the experimental protocols for preparing and testing the brain specimens. 
In Section \ref{expres}, we present the collected and filtered data and describe the filtering strategy adopted to keep meaningful experimental measurements. 
We then accurately model the data with the Mooney-Rivlin model in Section \ref{secmod} and obtain material parameters of brain matter which compare well with those found from other tests. To further validate the analytical modelling we implement Finite Element (FE) simulations in Abaqus to mimic the experiments and we finally use the estimated mechanical parameters to simulate a rotational head impact.\\
\indent Our main finding is that brain matter exhibits the normal Poynting effect, i.e. it tends to expand along its axis when twisted\cite{poynting1909pressure}. 
As noted by Rivlin\cite{rivlin1949large}, the Poynting effect is a nonlinear elastic effect \emph{par excellence} and cannot be explained by the linearised theory. 
It was present in all the cylindrical samples we tested, and when we simulated the twisting rapid motion of a head in a Finite Element model, we found that large vertical stresses developed in the whole brain also.


\section{Materials and methods}\label{matmet}


In this section, we give a brief description of the procedure for preparation and testing of the brain samples.


\subsection{Tissue preparation}\label{tissueprep}

\begin{figure}[h!]
\centering
\includegraphics[scale=0.38]{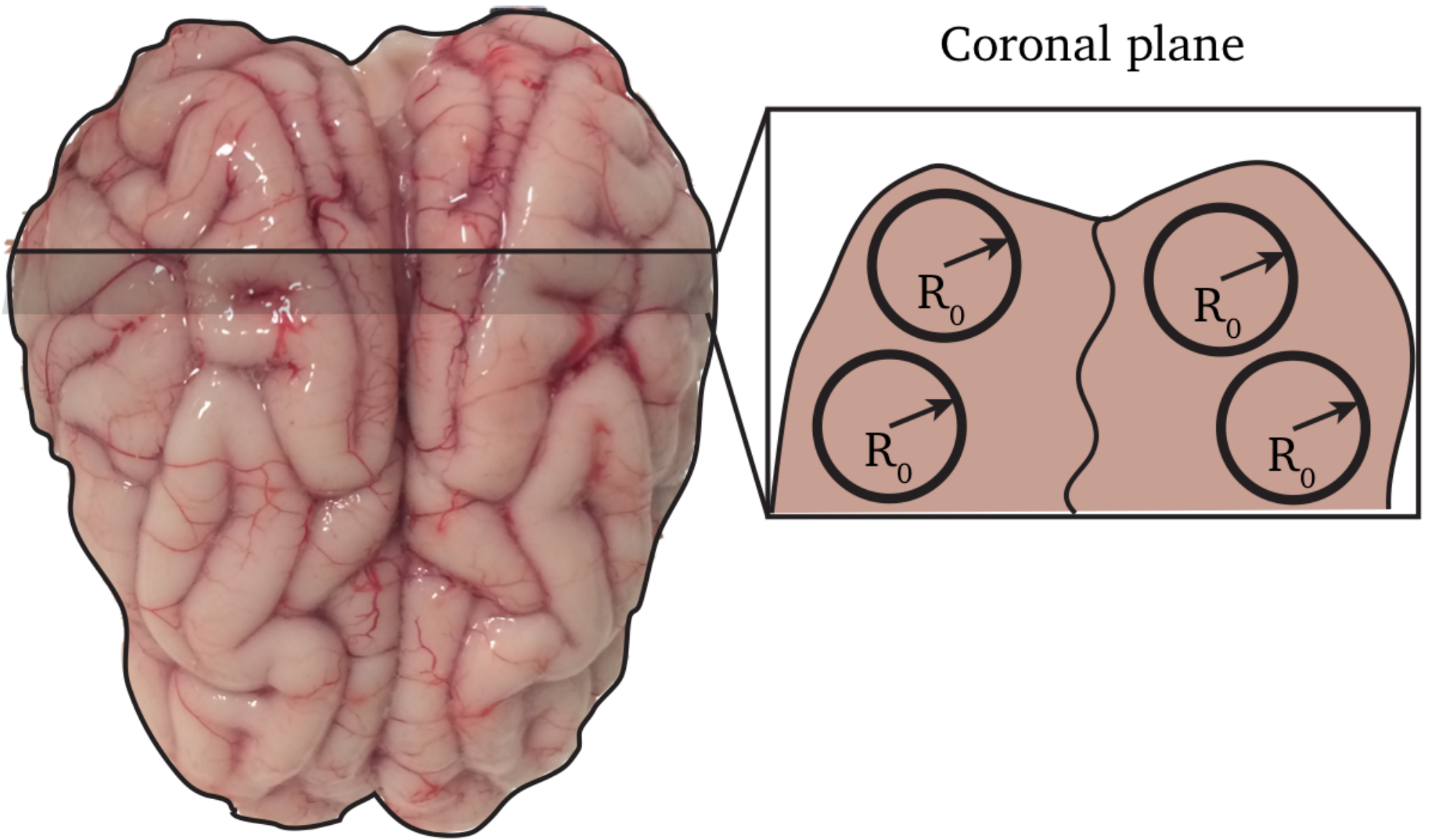}
\caption{Sketch of the cutting map of a fresh porcine brain (top view). The cylindrical samples were obtained by cutting a slice of brain excised from the coronal plane into cylinders of radius $R_0=\SI{10}{\milli\metre}$.}
\label{fig:samples}
\end{figure}

Six fresh porcine heads were obtained from a local abattoir from freshly killed 22 week old mixed sex pigs. The scalp was removed using a scalpel and the cranial bone was removed using an oscillating saw. Following removal of the skull, the meninges tissue was removed using surgical scissors. Finally, following resection of connective and vascular tissue and separation from the spinal cord, the undamaged brain was placed in PBS. The brains had an average (maximum) length, width and height of $7.5\pm0.3\,\si{\centi\metre}$, $6.4\pm0.3\,\si{\centi\metre}$ and $3\pm0.3\,\si{\centi\metre}$, respectively. A stainless steel cylindrical punch of $\SI{20}{\milli\metre}$ diameter was used to remove cylinders from each hemisphere as shown in Fig. \ref{fig:samples}. Each long cylindrical sample was then cut to samples of approximately $\SI{10}{\milli\metre}$ using a scalpel and template. The exact height of the specimen was measured again prior to testing. Mixed grey and white matter cylindrical samples were placed in PBS solution in multi-well plates of $\SI{20}{\milli\metre}$ diameter and placed in a fridge for less than $2$ hours while all samples were prepared. 

\subsection{Mechanical Testing}\label{mechtest}
A Discovery HR2 Hybrid Rheometer with parallel plates was used for all mechanical testing.  This device has a torque resolution of $\SI{0.1}{ \nano\newton\metre}$ and a normal force resolution of $\SI{0.5}{\milli\newton}$. To enable easy removal, masking tape was applied to both faces of the parallel plates, and then samples were glued to the tape using cyanoacrylate.  All testing was performed at room temperature and the samples were kept hydrated until the beginning of the test. A cylindrical Peltier plate with radius $r_p=\SI{10}{\milli\metre}$ was used. The rehometer was controlled through the TRIOS Software (v$4.3.1$). Each test consisted of a single \textit{stress growth} step for a duration of $\SI{10}{\second}$. 
The distance $H$ between the top plate and the bottom of the instrument was adjusted until the normal force was zero at the beginning of each test. 
The  twist rate (angular velocity of the upper plate per unit height) was  $\dot{\phi} =  \three$. 


\section{Experimental Results}
\label{expres}

\begin{figure}[t]
\centering
\subfigure[][Torque]{\includegraphics[scale=0.28]{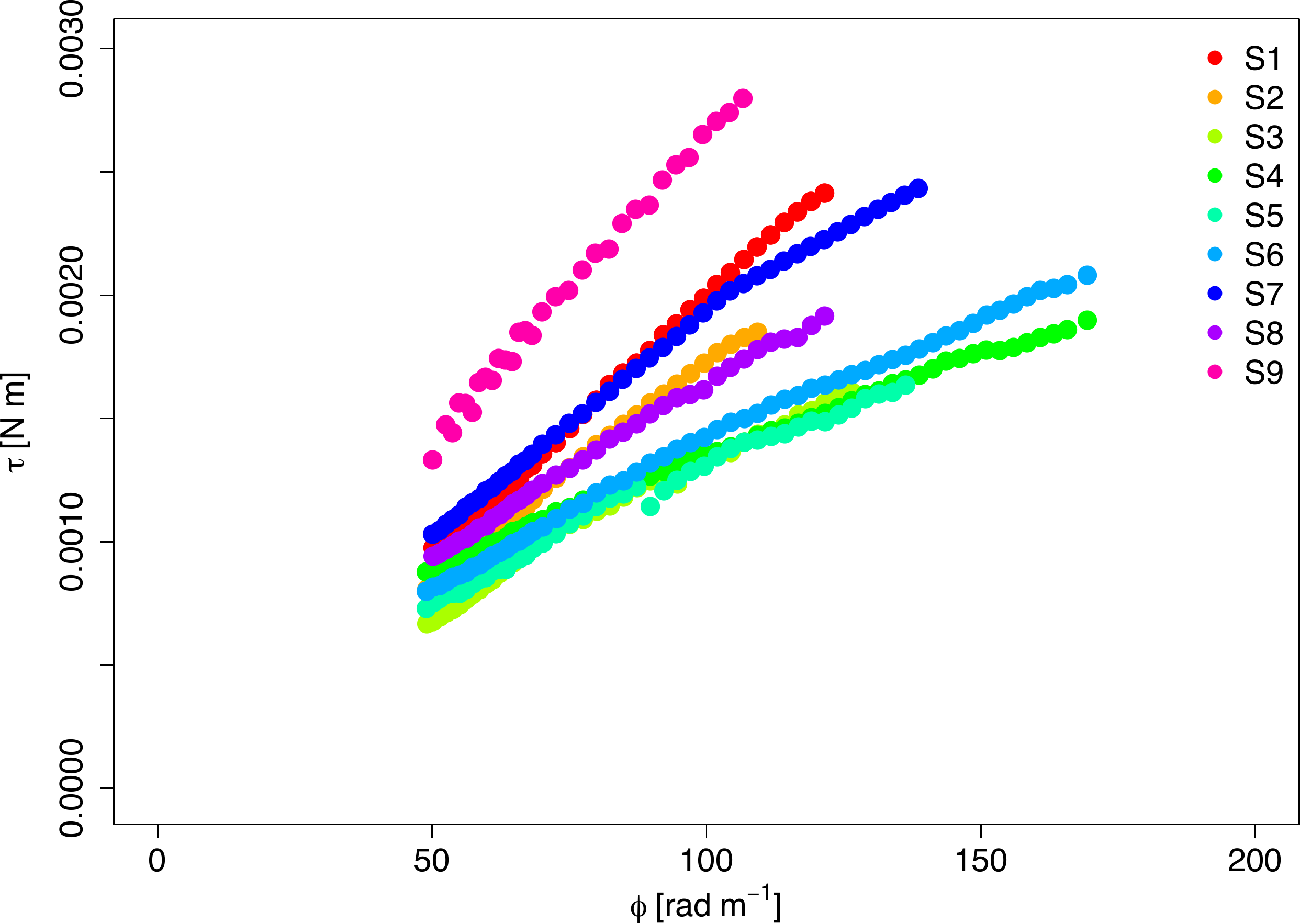}}
\subfigure[][Normal force]{\includegraphics[scale=0.28]{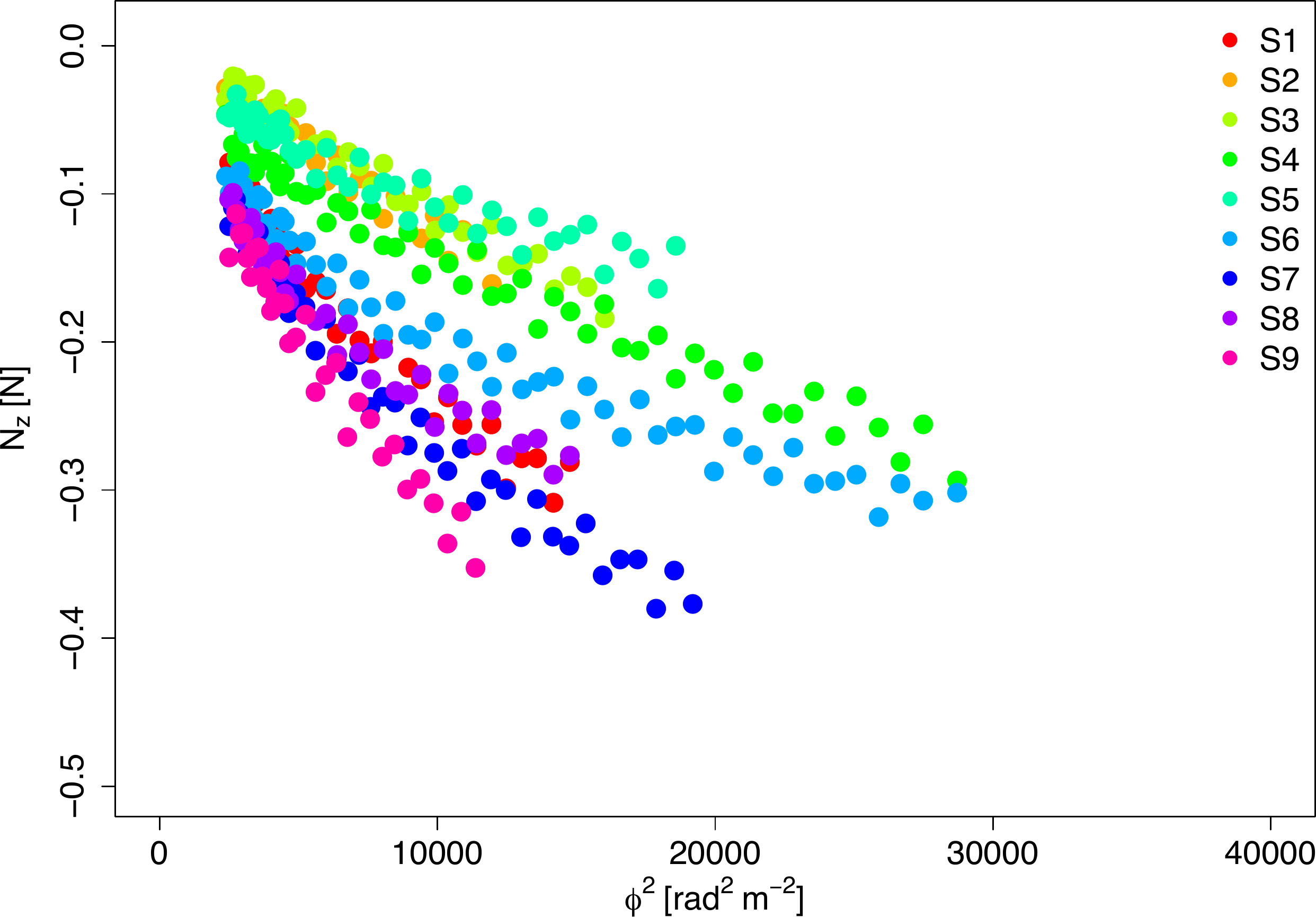}}\caption{Results of the torsion tests performed at a twist rate $\dot{\phi} =  \three$ on nine cylindrical samples of brain tissue excised from the coronal plane. (a) Torque $\tau$ vs twist $\phi$ and (b) normal force $N_z$ vs twist squared $\phi^2$, measured for cylindrical samples with initial radius $R_0 =\SI{10}{\milli\metre}$.}\label{fig:rawdata}
\end{figure}

In this section, we describe the filtering procedure required to get a clean set of data, ready for model fitting and parameter estimation. 
\\
\indent
\begin{figure}[b!]
\centering
\subfigure[][Representative torque and normal force output]{\includegraphics[scale=0.31]{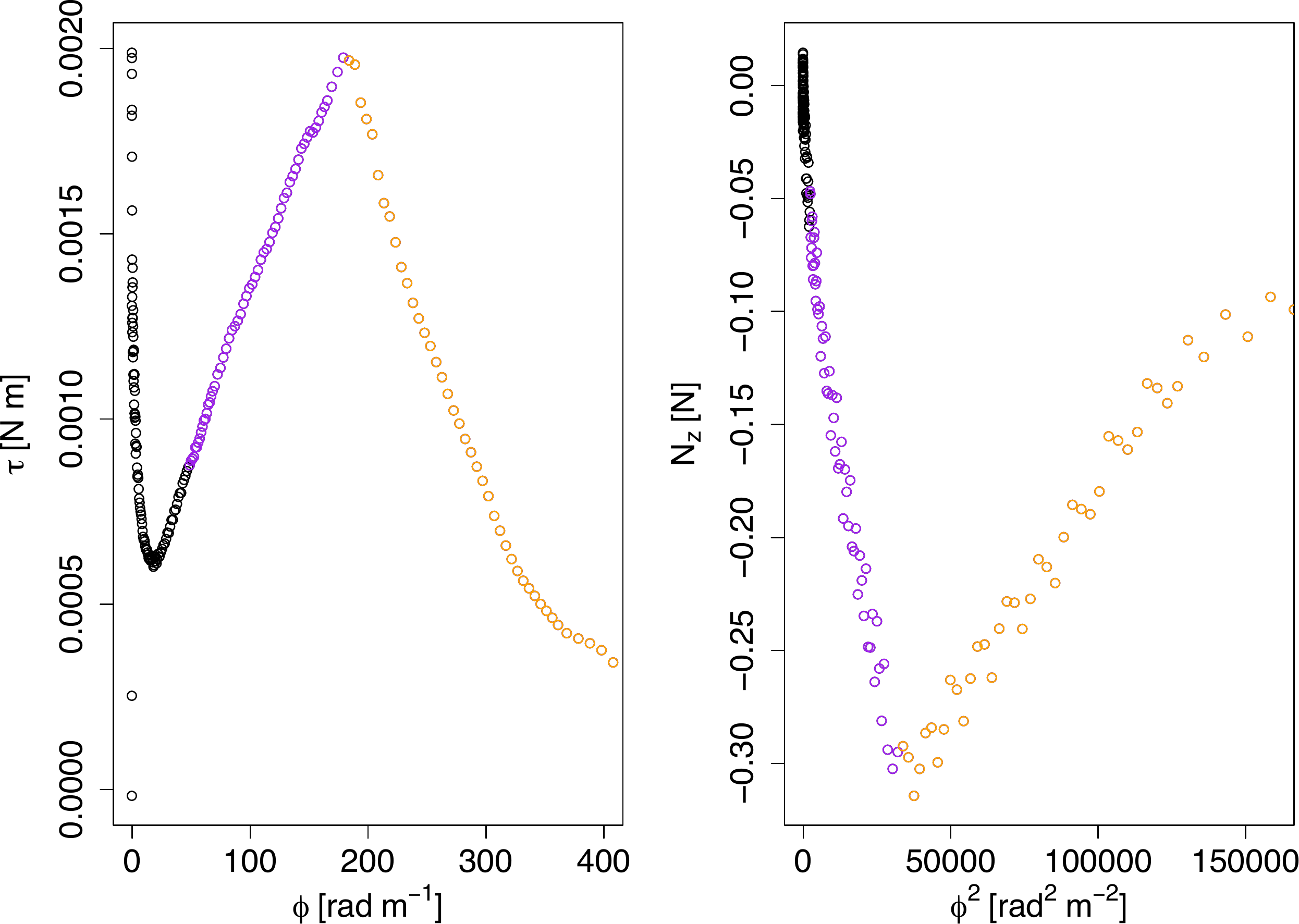}}
\subfigure[][Strain rate monitoring]{\includegraphics[scale=0.31]{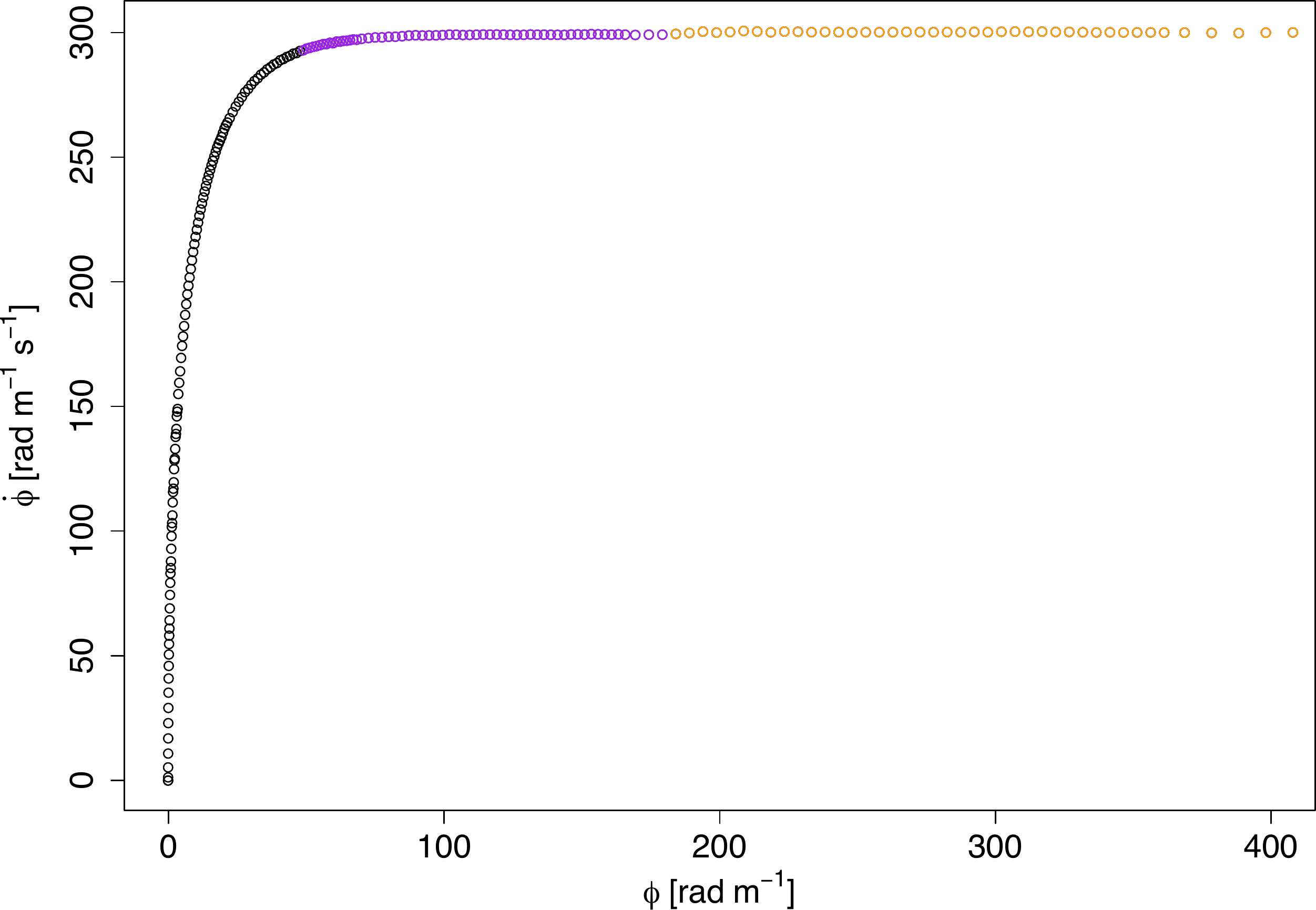}}
\subfigure[][Data output vs strain rate]{\includegraphics[scale=0.31]{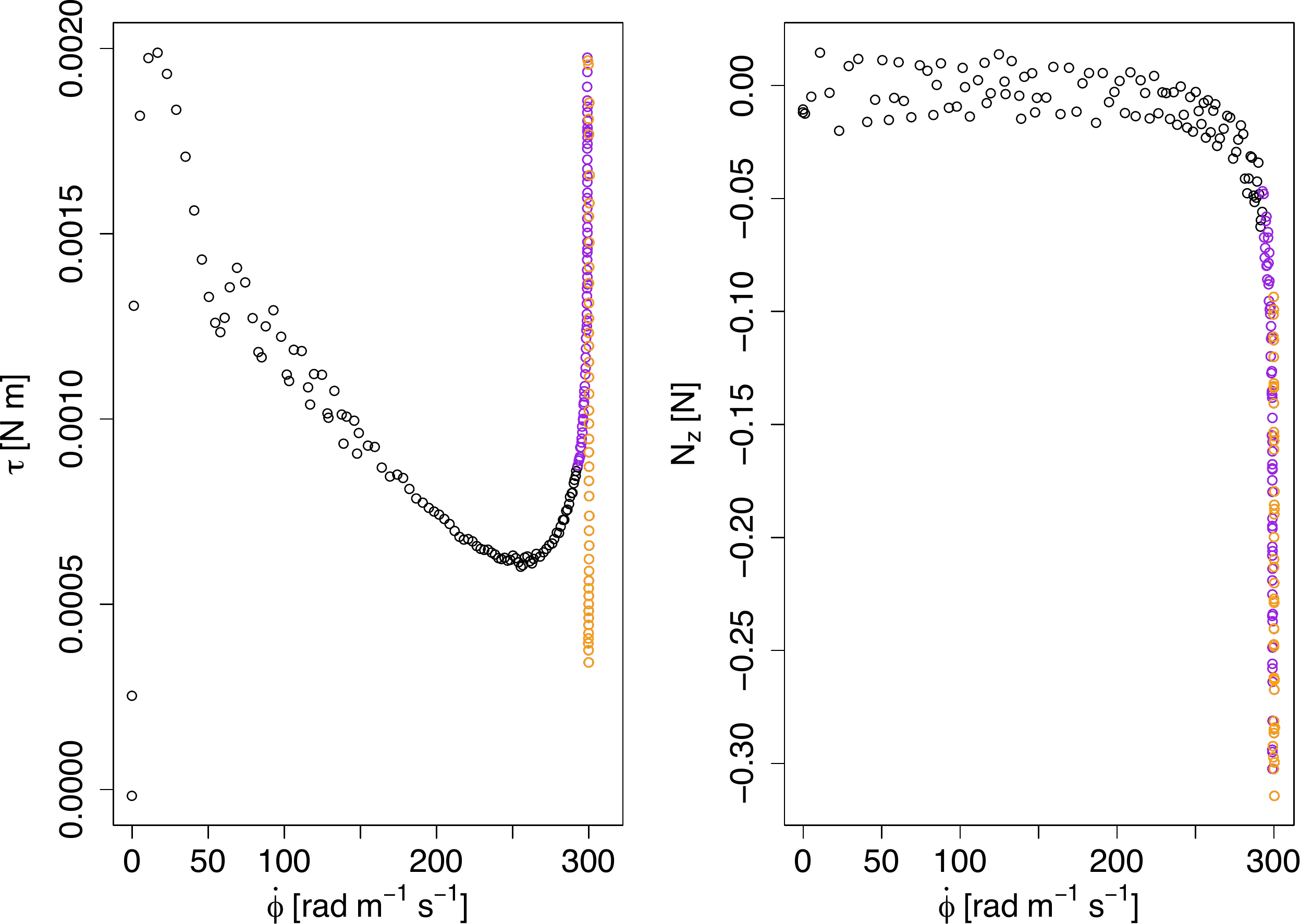}}\caption{Original collected data: (a, left) torque against twist and (a, right) normal force against twist squared; (b) twist rate against twist; (c, left) torque and (c, right) normal force against twist rate.
Black data: ramping towards the constant twist rate, see (b); Purple: proper data of torsion; Orange: data after breaking point, see (a). 
}\label{fig:filtering}
\end{figure}
Figure \ref{fig:filtering} shows a set of typical output data from the rheometer. 
In \ref{fig:filtering}(a) the torque $\tau$ and the normal force $N_z$ are plotted against the twist $\phi$ and the twist squared, respectively. 
The output twist $\phi$ in the plots is the angle of rotation $\alpha$ per unit length $\phi=\alpha/H$. 
From the data shown in Figure \ref{fig:filtering}(a), we identify three regions, for both the torque and the force data: (i) a noisy region (in black), at the very beginning of the test; (ii) a linear region (in purple) bounded by a maximum (minimum), and (iii) a decaying region (in orange) towards the end of the experiment. 
\\
\indent
The initial noisy region is due to the delay time of the instrument in reaching a constant twist rate. The plot in Figure \ref{fig:filtering}(b) shows the twist rate against the twist and clearly highlights the initial region where the upper plate is accelerating to reach the constant twist rate  $\dot{\phi} =  \three$. 
The plots in Figure \ref{fig:filtering} (c) further show that some data were actually generated at twist rate lower than $\three$. 
Only the data generated at  $\three$ was thus considered in the following analysis. 
\\
\indent
The other filtering criterion is the breaking point of the sample, which is identified clearly by a steep drop and rise in the plots of Figure \ref{fig:filtering} (a), indicating that an irreversible change in the mechanical response of the tissue occurred. 
%
Therefore, all data points after the breaking point were discarded. 
\\
\indent
The remaining ``good'' data obtained from nine samples $S_1, \ldots, S_9$ are shown in Figure \ref{fig:rawdata}.

%


\section{Modelling}\label{secmod}


To fit the experimental data and get a quantitative estimation of the behaviour of the brain in torsion, we first analyse the data, and then reproduce the mechanical tests theoretically. Finally we perform Finite Element simulations in Abaqus. 


\subsection{Theory}\label{sec:theo}


Here we calculate the torque $\tau$ and the normal force $N_z$ required to maintain a cylindrical sample of initial radius $R_0$ and initial height $L_0$ in a state of torsion.\par 
As mentioned in Section \ref{mechtest}, the normal force was set to zero before commencing each test. 
However, the force transducer of the HR2 rheometer has a sensitivity of $\SI{0.01}{\newton}$, so that variations of the force within that range are not detected by the instrument. 
We therefore expect that the sample undergoes a small contraction prior to the transducer picking up a meaningful value for the force.
Mathematically, we superpose an axial contraction to the actual rotation so that the total deformation is written in cylindrical coordinates as follows:
\begin{equation}\label{def}
r={R}/\sqrt{\lambda},\qquad \theta=\Theta+\phi \lambda Z,\qquad z=\lambda Z,
\end{equation}
where $\lambda$ is the  (tensile or compressive) pre-stretch, $\phi\!=\!{\alpha}/{(\lambda L_0)}$ is the twist per unit height  and $\alpha$ is the angle of rotation in radians. 
The corresponding deformation gradient $\textbf{F}$ is then:
\begin{equation}\label{defgrad}
\textbf{F}=\left(\begin{array}{ccc}
 {1}/{\sqrt{\lambda}} & 0 & 0 \\
 0 & {1}/{\sqrt{\lambda}} & r\phi\lambda \\
 0 & 0 & \lambda
\end{array}\right).
\end{equation}
\indent
We are interested in the elastic behaviour of brain matter, which we assume to be isotropic and incompressible.
In view of the linear dependence of the torque with respect to the twist highlighted by the results in Figure \ref{fig:rawdata}, we conclude that the constitutive behaviour of the brain must be modelled with a Mooney-Rivlin strain energy function\cite{mangan2016strain}:
\begin{equation}
W=c_1(I_1-3)+c_2(I_2-3),
\end{equation}
where $c_1$, $c_2$ are constants,  $I_1=\tr[\textbf{B}]$, $I_2 = \tr[\textbf{B}^{-1}]$ and $\textbf{B}=\textbf{F}\textbf{F}^{\T}$.
For this model the shear modulus is $\mu=2(c_1+c_2)$. 
The corresponding constitutive equation for the Cauchy stress $\boldsymbol{\sigma}$ reads:
\begin{equation}
\boldsymbol{\sigma} = 2 c_1\textbf{B} - 2 c_2\textbf{B}^{-1} - p \textbf{I},
\end{equation}
where $p$ is the Lagrange multiplier introduced to enforce incompressibility and $\textbf{I}$ is the identity matrix.
\\
\indent
The principal stretches are the square roots of the eigenvalues of $\textbf{B}$. 
The intermediate stretch $\lambda_1=1$  is associated with the radial direction, and the maximum and minimum stretches $\lambda_2$ and $\lambda_3$ are obtained by solving the following equations:
\begin{equation}
(\lambda_2\lambda_3)^2=\dfrac{1}{\lambda},\qquad \lambda_2^2 + \lambda_3^2 = 
 \dfrac{1}{\lambda} + \lambda^2 + (\lambda \phi r)^2.
\end{equation}
\indent 
The elastic equilibrium of the deformation is translated as the following problem:
\begin{equation}\label{equi}
\dfrac{\de}{\de r}\sigma_{rr}(r) + \dfrac{\sigma_{rr}(r)-\sigma_{\theta\theta}(r)}{r}=0,\qquad \sigma_{rr}(r_0)=0,
\end{equation}
with solution:
\begin{equation}\label{sigmasol}
\begin{split}
\sigma_{rr}(r)&= c_1 (r^2 - r_0^2) \lambda^2 \phi^2,\\
\sigma_{\theta\theta}(r)&=c_1 (3 r^2 - r_0^2) \lambda^2 \phi^2,\\
\sigma_{zz}(r)&=c_1\Big(2\, \dfrac{\lambda^3 -1}{\lambda} + (r^2 - r_0^2)\lambda^2\phi^2\Big)\\
&+2\,c_2\Big( \dfrac{\lambda^3 - 1}{\lambda^2} +  r^2 \lambda \phi^2\Big),\\
\sigma_{\theta z}(r)&=2  (c_2 + c_1 \lambda) r \lambda\phi.
\end{split}
\end{equation}
These formulas were first established by Rivlin\cite{rivlin1949large}, see Appendix A for details.\\
\indent
Now, the torque $\tau\!=\!2 \pi\int_0^{\nicefrac{R_0}{\sqrt{\lambda}}}r^2 \sigma_{\theta z}(r)\de r$ and the normal force $N_z\!=\!2 \pi\int_0^{\nicefrac{R_0}{\sqrt{\lambda}}}r\, \sigma_{z z}(r)\de r$ that have to be applied to the cylinder to maintain the deformation in \eqref{def} are:
\begin{equation}\label{tau}
\tau= \pi \, R_0^4\Big( c_1+ \dfrac{c_2}{\lambda}\Big) \phi=\mathcal{A} \phi 
\end{equation}
\begin{equation}\label{Nz}
\begin{split}
N_z&= -2 \pi\, R_0^2\Big(c_1+\dfrac{c_2}{\lambda}\Big)\Big(\dfrac{1-\lambda^3}{\lambda^2}\Big)-\pi\,R_0^4\Big(\dfrac{c_1}{2}+\dfrac{c_2}{\lambda}\Big)\phi^2\\
&=\mathcal{C}+\mathcal{B}\phi^2
\end{split}
\end{equation}
where the constants $\mathcal{A},\mathcal{B},\mathcal{C}$ introduced above and the parameters $c_1,c_2$ are linked by the following relations:
\begin{align}
c_1=2\,\dfrac{\mathcal{A}+\mathcal{B}}{\pi R_0^4}, \qquad 
\dfrac{c_2}{\lambda}=-\dfrac{\mathcal{A}+2\mathcal{B}}{\pi R_0^4},\label{c1c2}
\end{align}
and the pre-stretch $\lambda$ is the unique real and positive root of the following cubic:
\begin{equation} \label{cubic}
2 \mathcal{A}(\lambda^3 - 1) -\mathcal{C} R_0^2 \lambda^2  = 0.
\end{equation}
Note, as expected for the Mooney-Rivlin model, the linear dependence of the torque on the twist and of the normal force on the twist squared, see \eqref{tau} and \eqref{Nz}. 
The coefficient $\mathcal{B}$ is associated with the Poynting effect displayed by the sample and is due almost entirely to the twist, whereas the coefficient $\mathcal{C}$ accounts for the pre-stretch only. 
When $\lambda=1$, i.e. in \emph{pure torsion}, we have $\mathcal A = \pi R_0^4(c_1+c_2)$, $\mathcal{B} = -\pi R_0^4(c_1/2 +c_2)$, $\mathcal{C}=0$. 
The values of $\lambda$ in Table \ref{tab:resultsgeom} show that samples $S_2$ and $S_3$ experience less than 1\% pre-stretch; hence for those two samples, $\mathcal{B}$ provides an effective measure of the exact Poynting effect, i.e. in the absence of a normal compressive force, the samples would expand axially.


\subsection{Parameters estimation}


To fit the data in Figure \ref{fig:rawdata}, we use the open-source software RStudio (version $1$.$1$.$383$). 
The function \texttt{lm} (from the package \textit{stats}) allows to perform a linear regression on the data sets $\{\phi,\tau\}$ and $\{\phi^2,N_z\}$. By calling the function \texttt{lm} on the set $\{\phi,\tau\}$ we obtain the coefficient $\mathcal{A}$ from equation \eqref{tau} and the linear regression on $\{\phi^2,N_z\}$ gives us the coefficients $\mathcal{B},\mathcal{C}$ appearing in equation \eqref{Nz}. 

\begin{table}[b!]\centering
\begin{tabular}{|c|c|c|c|c|}
\hline 
sample & $\mu [\si{\pascal}]$ & $c_2 [\si{\pascal}]$ & \multicolumn{1}{c|}{$R_{\tau}^2$} & \multicolumn{1}{c|}{$R_{N_z}^2$}  \\ \hline 
$S_1$ & $1232.50$ & $294.45$ & 0.999 & 0.946 \\ \hline 
$S_2$ & $1092.31$ & $235.84$ & 0.998 & 0.939 \\ \hline 
$S_3$ & $766.95$ & $310.60$ & 0.988 & 0.966 \\ \hline 
$S_4$ & $491.14$ & $201.22$ & 0.996 & 0.958 \\ \hline 
$S_5$ & $656.96$ & $59.68$ & 0.988 & 0.87 \\ \hline 
$S_6$ & $644.59$ & $113.75$ & 0.994 & 0.92 \\ \hline 
$S_7$ & $952.36$ & $347.26$ & 0.993 & 0.947 \\ \hline 
$S_8$ & $803.12$ & $401.41$ & 0.997 & 0.925 \\ \hline 
$S_9$ & $1460.17$ & $710.29$ & 0.995 & 0.962 \\ \hline 
mean$\pm$SD & $900\pm 312$ & $297\pm 189$ \\\cline{1-3}
\end{tabular}\caption{Estimated elastic parameters: the shear modulus $\mu\!=\!2(c_1\!+\!c_2)$, the Mooney-Rivlin parameter $c_2$ and the coefficient of determination $R^2$. Mean values with standard deviation for $\mu$ and $c_2$ are also calculated in the last row.}\label{tab:resultsfit}
\end{table}

\begin{table}[b!]\centering
\begin{tabular}{|c|c|c|c|c|}
\hline 
sample & $\lambda$ & $l_0 \,[\si{\milli\metre}]$ & $r_0 \,[\si{\milli\metre}]$ & $\lambda_2^{\text{max}}$\\ \hline 
$S_1$ & $0.93$ & $12.62$ & $10.36$ & $1.82$ \\ \hline 
$S_2$ & $0.99$ & $16.02$ & $10.05$ & $1.73$ \\ \hline 
$S_3$ & $0.99$ & $15.92$ & $10.05$ & $1.81$ \\ \hline 
$S_4$ & $0.89$ & $12.95$ & $10.59$ & $2.14$ \\ \hline 
$S_5$ & $0.94$ & $13.26$ & $10.31$ & $1.85$ \\ \hline 
$S_6$ & $0.85$ & $14.19$ & $10.84$ & $2.04$ \\ \hline 
$S_7$ & $0.89$ & $9.51$   & $10.59$ & $1.87$ \\ \hline 
$S_8$ & $0.89$ & $12.22$ & $10.59$ & $1.68$ \\ \hline 
$S_9$ & $0.95$ & $10.14$ & $10.26$ & $1.69$ \\ \hline 
\end{tabular}\caption{Geometry of the samples after pre-compression, prior to twisting: the estimated axial stretch $\lambda$, the length $l_0=\lambda L_0$ (measured by the instrument), the radius $r_0=\nicefrac{1}{\sqrt{\lambda}}R_0$ of the nine samples and the maximum value of the greatest principal stretch $\lambda_2$ before sample breaking.}\label{tab:resultsgeom}
\end{table}

Moreover, the fit on $\{\phi^2,N_z\}$ uses a weighted (with respect to $\phi^2$) least squares method. 
Finally, we input the coefficients $\mathcal{A},\mathcal{B}$ and $\mathcal{C}$ into equations \eqref{c1c2} to get the elastic parameters $c_1$ and $c_2$. 
The results of the linear regression are shown in Table \ref{tab:resultsfit}. The mean values for the elastic parameters are $\mu=\SI{900}{\pascal}$ and $c_2=\SI{297}{\pascal}$, respectively.  The dimensions of the nine samples after the compression are summarised in Table \ref{tab:resultsgeom}, where the length $l_0$ and radius $r_0$ are given for each sample, as well as the pre-stretch computed from \eqref{cubic}and the maximum values of the highest principal stretch $\lambda_2^\text{max}$, attained at breaking point on the periphery of the top face.


\subsection{Computational validation}


We performed brain torsion simulations using ABAQUS Standard $6.14$-$1$ to validate our analytical modelling of the deformation. 
The initial cylindrical geometry of the sample was obtained by setting the radius $R_0=\SI{10}{\milli\meter}$ and the height $L_0={l_0}/{\lambda}$, calculated according to Table \ref{tab:resultsgeom}. 
We used a mesh of $78,750$ hexahedral elements (C3D8) with hybrid formulation to reproduce exact incompressibility and we assigned a Mooney Rivlin model with material parameters in Table \ref{tab:resultsfit} to account for hyperelasticity. \par
To simulate the twist, we first defined a reference point at the centre of the top surface of the cylinder, which we then coupled with all points on the surface, and finally we assigned a rotational displacement around the longitudinal axis (ramp form of amplitude $\SI{3}{\radian}$) whilst setting the other degrees of freedom to zero. The bottom surface of the cylinder was encastred. The output variables were the resultant axial force (RF3) and torque (RM3).\par 
To perform the simulations we chose two specimens: $S_1$ and $S_2$. An additional step, prior to the torsion, was added to simulate the 7\% pre-compression (see Table \ref{tab:resultsgeom}) undergone by $S_1$. 
\begin{figure}[b!]
\centering
\subfigure[][Torque]{\includegraphics[scale=0.44]{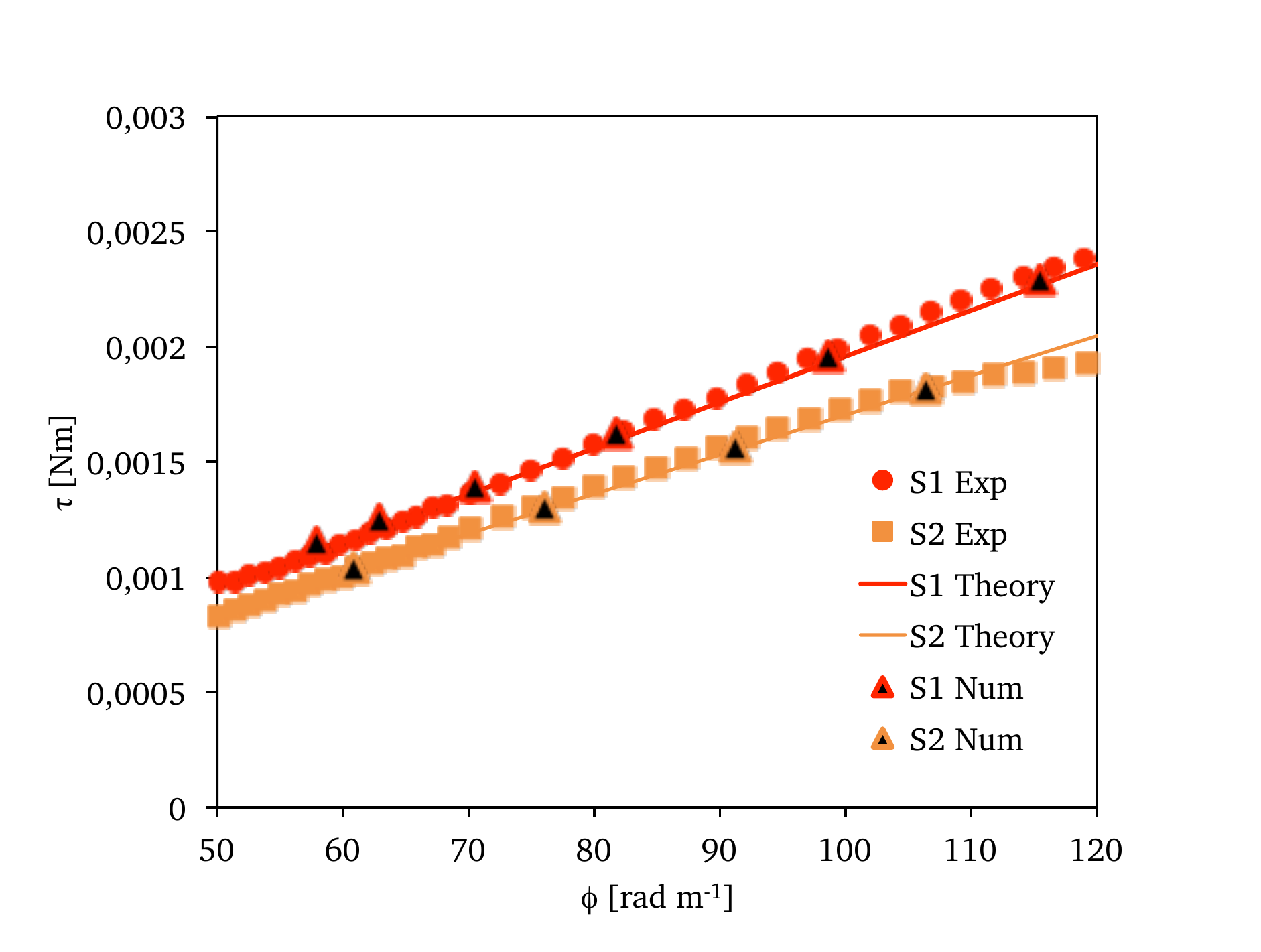}}
\subfigure[][Normal force]{\includegraphics[scale=0.44]{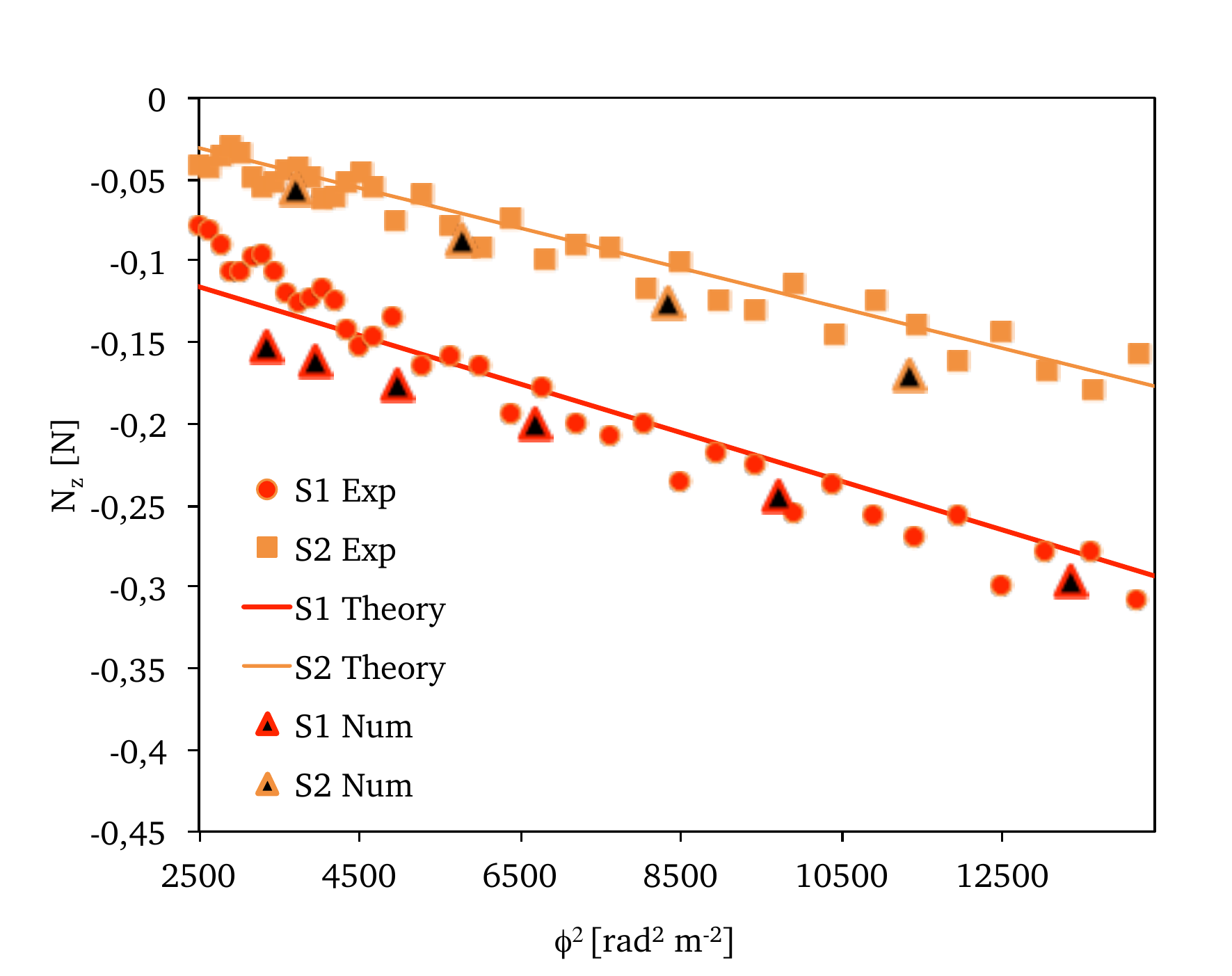}}
\caption{Comparison of the resultant torque $\tau$ and normal force $N_z$ for $S_1$ and $S_2$. Results of the numerical simulations in Abaqus (triangles), analytical predictions with the models \eqref{tau} and \eqref{Nz} (solid lines) and experimental data (red circles for $S_1$ and orange squares for $S_2$).}\label{fig:resnum}
\end{figure}
The results are shown in Fig. \ref{fig:resnum}. We see that the numerical simulations validate the predictions of the analytical model described in Section \ref{sec:theo}. The torque and the normal force calculated in Abaqus are consistent with the analytical predictions and the measured data for both cases with and without pre-compression ($S_1$ and $S_2$, respectively). 
We note that there is a small mismatch between the analytical and the numerical normal force for $S_1$. 
This is due to the longitudinal bulging of the sample occurring during the 7\% compression phase, which results in a non-homogeneous deformation along the axis of the cylinder. 


\section{Discussion and conclusions}


Now we compare the results obtained here for torsion tests with those obtained elsewhere for simple shear tests and for torsion modelled as simple shear.

We begin by recalling that the deformation gradient for uni-axial compression in the $Z$ direction, followed by simple shear of amount $\kappa$ in the $YZ$ plane, has the form\cite{rajagopal1987new}
\begin{equation}
\textbf{F}=\left(\begin{array}{ccc}
 1/\sqrt \lambda & 0 & 0 \\
 0 & 1/\sqrt \lambda & \lambda \kappa \\
 0 & 0 & \lambda\end{array}\right).
\end{equation}
Hence, we see from comparison with \eqref{defgrad} that there is a formal connection between torsion and simple shear. 
However, the equivalence is \emph{local} only, as simple shear is homogeneous but torsion is not: the amount of ``shear'' experienced by an element in torsion ($\kappa = r \phi = r \alpha/H$) depends on the dimensions of the sample and the position of the element.
Thus, it does not make sense to compare the amount of shear and the shear rate experienced by all elements in a simple shear experiment with the amount of ``shear''  and the ``shear'' rate experienced by a given element at a given location for a given sample dimension in a torsion experiment.
Despite this disconnect, finite shear and torsion are often confused in the literature, and torsion experiments in rheometers are routinely modelled as simple shear, see for example the papers cited in the extensive review by Chatelin et al.\cite{chatelin2010fifty}.\par
Here we note that the results presented in Table \ref{tab:resultsfit} show that cylindrical samples of brain matter with initial radius $R_0=\SI{10}{\milli\metre}$ twisted at a twist rate $\dot{\phi}=\three$ behave as Mooney-Rivlin materials with a shear modulus $\mu=900\pm312\si{\pascal}$. 
This value is in the same range of values found by Rashid et al.\cite{rashid2013mechanical} when a block of porcine brain matter was sheared at a shear rate of $\dot \kappa=\SI{30}{\second^{-1}}$, estimated to be conducive to diffuse axonal injury\cite{morrison2003tissue,morrison2006vitro}. 
Moreover, from Table \ref{tab:resultsgeom} we note that the values of the principal stretch (greatest stretch) $\lambda_2$ at the breaking point are in the range $1.67-2.14$ which corresponds to extensions of 67\% to 114\%. 
These values of strain are well above the estimate axonal strain thresholds associated to diffuse axonal injury ($>0.05$) \cite{yap2017mild} and to white matter damage in the optical nerve ($>0.34$)\cite{bain2000tissue}. 
 
In addition, here we are able to directly estimate the second Mooney-Rivlin coefficient $c_2$ from the normal force data and thus provide a direct measure of the Poynting effect. 
This quantity cannot be measured from shear stress data alone, although its sign (positive for porcine brain matter) can be deduced by piercing a hole in one of the plattens\cite{destrade2015extreme}.
We note also that in contrast to simple shear, where a neo-Hookean material ($c_2=0$) does not display the Poynting effect\cite{destrade2015extreme},  the same material does have a non-zero normal force in torsion\cite{horgan2015reverse}. \par
Our conclusion is that brain matter exhibits a real and large Poynting effect in torsion, which is bound to lead to the development of large normal forces in an impacted brain. 

\begin{figure}[t!]
\centering
\includegraphics[width=0.48\textwidth]{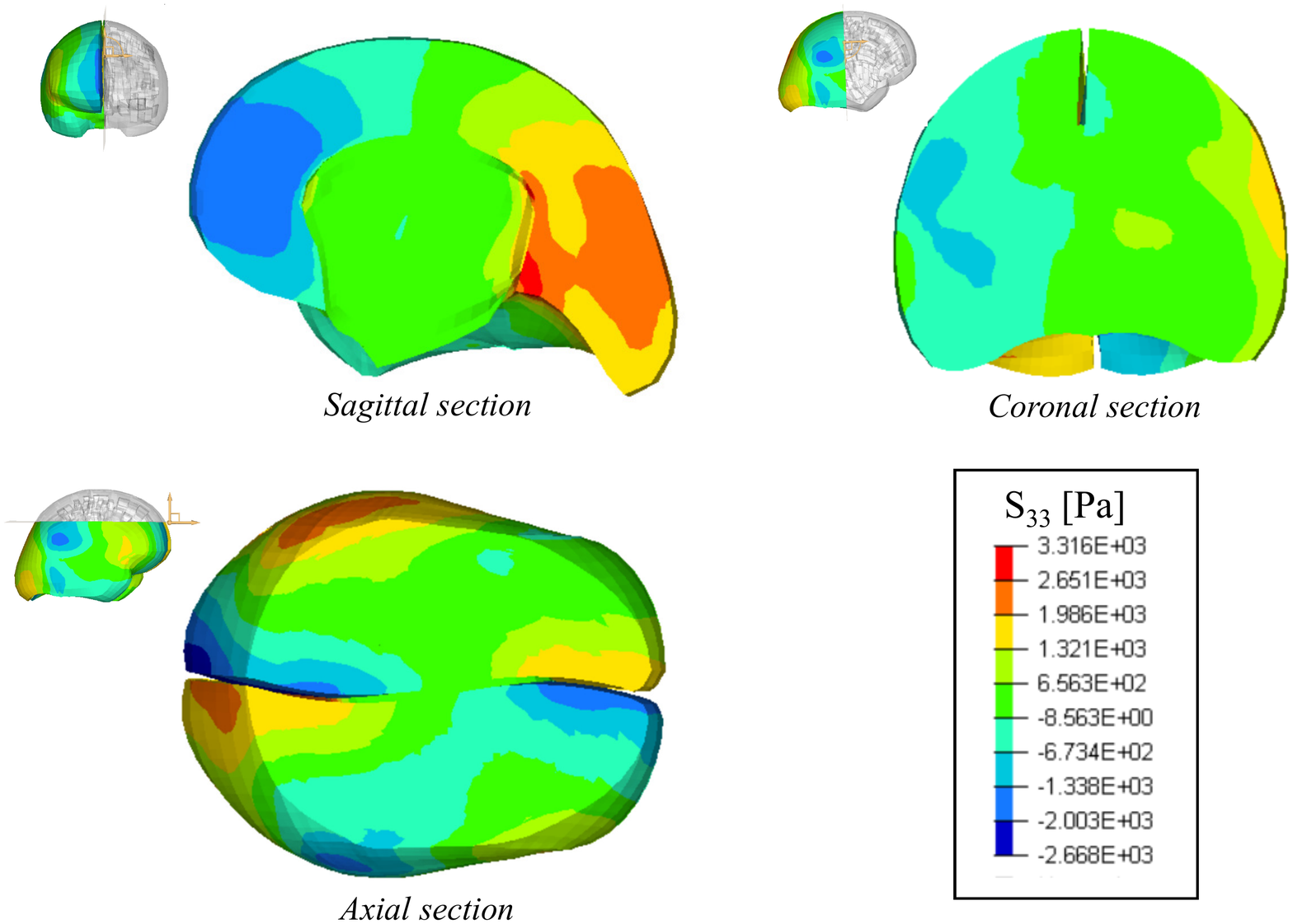}\caption{Results of the FE simulations of a rotational head impact, performed with the UCDBT Model. Distribution of the stress component $S_{33}$ across the \textit{Sagittal}, \textit{Coronal} and \textit{Axial} planes.}\label{fig:head}
\end{figure}

In order to investigate the existence and magnitude of axial forces during twisting head impacts, we used Abaqus/Explicit to simulate a rotational impact with the University College Dublin Brain Trauma Model (UCDBTM) developed by Horgan and Gilchrist\cite{horgan2003creation}. 

We applied a rotational acceleration in the axial plane to the centre of gravity of the head, peaking at $\SI{2,170}{\,\radian\, \second^{-2}}$.
This value of rotational acceleration is in the range of accelerations experienced in boxing \cite{walilko2005biomechanics}.
We used the mean values of the estimated parameters $\mu$ and $c_2$ in Table \ref{tab:resultsfit} to model the mechanical behaviour of the brain matter.
In Fig \ref{fig:head}, the distribution of the axial stress component $S_{33}$ (where $\textbf{S}$ is the deviatoric part of the Cauchy stress) throughout the brain is shown for a rotational angle of $\alpha=\SI{0.52}{\radian}$. 
The three sections in the sagittal, coronal and axial planes, respectively, highlight areas where the stress reaches peaks of magnitude in the thousands of $\si{\pascal}$. 
Therefore the high normal stresses developing during rotational impacts could potentially contribute to Traumatic Brain Injury (TBI) during this type of impact. 
The results presented in this work thus open the path towards further studies to quantify the role played by normal forces in TBI, in particular with a view to define more accurate threshold criteria for TBI.


\section*{Appendix A}

By integrating the first of \eqref{equi} together with the initial condition we obtain:
\begin{equation}\label{equi1s}
\begin{split}
\sigma_{rr}(r)&=\int_r^{r_0}\dfrac{\Sigma_r-\Sigma_{\theta}(r)}{r}\de r\\
\sigma_{\theta\theta}(r)&=\Sigma_{\theta}(r)-\Sigma_r+\int_r^{r_0}\dfrac{\Sigma_r-\Sigma_{\theta}(r)}{r}\de r\\
\sigma_{zz}(r)&=\Sigma_{z}(r)-\Sigma_r+\int_r^{r_0}\dfrac{\Sigma_r-\Sigma_{\theta}(r)}{r}\de r\\
\sigma_{\theta z}(r)&=c_1B_{\theta z}(r)-c_2 \dfrac{B_{\theta z}(r)}{\Sigma}
\end{split}
\end{equation}
where
\begin{equation}\label{equi2s}
\begin{split}
&\Sigma_{r}-\Sigma_{\theta}(r)=2c_1\left(B_{rr}-B_{\theta\theta}(r)\right)+2c_2\left(\dfrac{B_{zz}}{\Sigma}-\dfrac{1}{B_{rr}}\right)\\
&\Sigma_{z}(r)-\Sigma_{r}=2c_1\left(B_{zz}-B_{rr}\right)+2c_2\left(\dfrac{1}{B_{rr}}-\dfrac{B_{\theta\theta}(r)}{ \Sigma}\right)\\
&\text{and }\quad \Sigma=B_{zz} B_{\theta\theta}(r)-B_{\theta z}^2(r)=1/B_{rr}
\end{split}
\end{equation}
Then, by substituting \eqref{defgrad} into \eqref{equi2s} and then into \eqref{equi1s} we obtain \eqref{sigmasol}.
\section*{Conflicts of interest}
There are no conflicts to declare.

\section*{Acknowledgements}
We thank Badar Rashid for Figures \ref{fig:protocols}(a), (b), (c); David McManus for help with the dissection of pig heads; Xiaolin Li for technical assistance with the rheometer; Christiane G\"orgen for help with fitting in RStudio and Giuseppe Saccomandi for insightful discussions on the modelling of torsion.
The work has received funding from the European Union's Horizon 2020 Research and Innovation Programme under the Marie Sk\l{}odowska-Curie grant agreement  No.705532 (Valentina Balbi and Michel Destrade).

\bibliography{Poynting_brain_arXiv.bbl} 
\bibliographystyle{vancouver} 

\end{document}